\documentstyle[preprint,aps]{revtex}
\begin{document}
\title{Re-entrant spin glass  and Magnetoresistance in
Co$_{0.2}$Zn$_{0.8}$Fe$_{1.6}$Ti$_{0.4}$O$_4$ spinel oxide}
\author{R. N. Bhowmik\footnote{e-mail address:rnb@cmp.saha.ernet.in} and R. Ranganathan}
\address{Experimental condensed Matter Physics Division,\\ 
Saha Institute of Nuclear physics,\\
 1/AF, Bidhannagar, Calcutta 700064, India}
\maketitle
\begin{abstract}
We have investigated the static and dynamical response of magnetic clusters in 
Co$_{0.2}$Zn$_{0.8}$Fe$_{1.6}$Ti$_{0.4}$O$_4$ spinel oxide, where
a sequence of magnetic phase transitions, {\it i.e}, paramagnetic(PM) to
ferromagnetic (FM) state at T$_C$ $\leq$ 270K and ferromagnetic to canted 
spin glass (CSG) state at T$_f$ $\leq$ 125K is observed. The time
dependence of remanent magnetization shows a non-equilibrium spin dynamics in 
the CSG state and above 130K an weak time dependent relaxation characterizes 
a canted ferromagnetic state which is followed up by no relaxation effect 
in the paramagnetic state. The field dependence of the magnetization confirms 
the absence of long range ferromagnetic order in the system. This brings the
idea that all the spins are not necessarily be infinite ordered inside the
clusters due to spin canting effects. 
The variation of the ferromagnetic and 
antiferromagnetic components and magnetic disorder inside the clusters 
shows some interesting magnetic and electrical properties in the system, {\it
viz}, field induced transition in M vs H data, re-entrant magnetic transition
in ac susceptiblity vs T data and re-entrant semi-conducting behaviour in resistivity vs T data.
\end{abstract}
\section{Introduction}
Recently, extensive research activities are going on a variety of colossal 
magneto resistance (CMR) materials, which includes substituted manganite
R$_{1-x}$A$_x$MnO$_3$ [R= trivalent rare earth ion, A= divalent alkaline rare 
earth ion] \cite{Mathur} and spinel ferrites such as 
Fe$_{1-x}$T$_x$CrS$_4$ [T=Cu,Zn etc.] \cite{Cava,Wang} and
Co$_x$Mn$_{3-y}$O$_4$ \cite{Kutty}. The substitution by magnetic or
non-magnetic ions has shown a tremendous effect in controlling the magnetic
and electrical properties of a magnetic materials. As for example, the parent 
manganite LaMnO$_3$ is an antiferromagnet and insulator. But the substituted
manganites show ferromagnetism with metallic behaviour below the Curie
temperature T$_C$ \cite{Mathur}. The competition between ferromagnetic 
(metallic) and antiferromagnetic (insulator) exchange interactions change the 
the polaronic hopping between
Mn$^{4+}$(t$^{3}_{2g}$e$_g$)-O-Mn$^{3+}$(t$^{3}_{2g}$e$^{1}_{g}$) in
substituted manganites \cite{Mahen}.
Besides the polaronic hopping, it has also been suggested that double 
exchange mechanism via Mn$^{3+}$-O-Mn$^{4+}$ ions, Jahn-Teller distortions due 
to Mn$^{3+}$ ions and strong electron-phonon coupling are   
contributing to CMR effect in perovskites materials \cite{Mathur}.\\ 
In general, there is no mixed valence Mn$^{3+}$ - Mn$^{4+}$ ions or double
exchange mechanism in spinel ferrites. In spinel lattices,
the anions (O$^{2-}$, S$^{2-}$ ions) form a cubic close packing, in
which the interstices are occupied by tetrahedral (form A sites or sublattice) 
and octahedral (form B sites or sublattice) coordinated cations, gives rise
the formula unit AB$_2$X$_4$ (A represent A site cations, B represent B site
cations, X represent anions). The competition between ferromagnetic and
antiferromagnetic superexchange interactions occurs between the spins of
inter-sublattices and intra-sublattices as A-O-B (J$_{AB}$ = inter-sublattice 
exchange interaction), B-O-B (J$_{BB}$ = B site intra-sublattice interaction) 
and A-O-A (J$_{AA}$ = A site intra-exchange interaction). 
In collinear spinel structure ${\vert}J_{AB}{\vert}>> {\vert}J_{BB}{\vert}>>
{\vert}J_{AA}{\vert}$ and the system shows long range ferrimagnetic 
(ferromagnetic) order \cite{Dorman}. 
If A sublattice magnetic dilution is below the percolation limit 
(C$_{A}$$\approx$ 0.33 -0.4), the J$_{AB}$ and J$_{BB}$
will be comparable and magnetic frustration appears in the B sublattice. 
Then B site spins form finite size clusters. The spins inside the clusters 
may be canted due to short range antiferromagnetic interactions of
nearest-neighbours.
The longitudinal components (S$_z$) of a canted spin will contribute long
range ferromagnetic ordering along the broken 
axis of symmetry (z axis) whereas the transverse 
components S$_t$ will contribute spin glass ordering in x-y plane
 \cite{Dorman}. 
The competition between ferromagnetic and antiferromagnetic interactions and
magnetic disorder in B sites have shown spin glass or re-entrant spin glass 
behaviour in spinel oxides \cite{Fiorani,Brand,Sing}.\\
Recently, it has been observed that the existence of ferromagnetic
clusters (magnetic polarons) with itinerent charges (charges are itinerant 
within the
clusters) in either an antiferromagnetic insulating matrix or paramagnetic
insulating matrix exhibits an unusual magnetic properties with large CMR 
effect \cite{Mathur,Mahen}.
It has been found theoretical by 
Aharoni et al. \cite{Aha1} and experimental by H\"{u}cker et al.
 \cite{Huck} on La$_{2-x}$Sr$_x$Cu$_{1-z}$Zn$_z$O$_4$ that 
charge compensating defects like magnetic ion vacancy
can significantly modify the effective exchange interactions by 
renormalizing the concentration of frustrated bonds and hopping conductivity 
of a system. The frustrating bond will behave like a magnetic dipole, which
makes the superexchange interactions more ferromagnetic inside the clusters.
It has also been suggested \cite{Sugi} that the existence of different 
valence cations in B site can enhance the conductivity in spinel oxides.
To test the effect of cation vacancy and existence of mixed
valence cations like Fe$^{2+}$/Fe$^{3+}$, extensive studies have been
performed in Fe$_{3-x}$M$_x$O$_4$ [M = Ti$^{4+}$,Zn$^{2+}$] spinel oxides
 \cite{Millot}. The results showed significant enhancement in conductivity with
increasing cation vacancy and mixed valence cations in B sites.\\ 
In order to understand the magnetic ordering of B site clusters with
intra spin canting, we have studied the Co$_{0.2}$Zn$_{0.8}$Fe$_2$O$_4$ spinel 
oxide \cite{rnbzn8}. We have considered that the magnetic clusters are
randomly distributed in the antiferromagnetic B site matrix.
In this presentation, we have replaced Fe$^{3+}$ by non-magnetic 
Ti$^{4+}$ in Co$_{0.2}$Zn$_{0.8}$Fe$_2$O$_4$ with the motivation to study the 
effects of Fe$^{2+}$/Fe$^{3+}$ mixture, cation vacancies in B sites. 

\section{Experimental}
\subsection{Sample preparation and Characterization}
We have prepared Co$_{0.2}$Zn$_{0.8}$Fe$_{1.6}$Ti$_{0.4}$O$_4$ solid solution 
by conventional solid state method. The stoichiometric amount of powder 
oxides of 99.5\% Co$_3$O$_4$ (from Fluka), 99.998\% Fe$_2$O$_3$ (from Johnson 
Matthey), 99.998\% ZnO (from Johnson Matthey) and 99.995\% TiO$_2$ (from 
Johnson Matthey) have been mixed and grinded for $\approx$ 2 hours. The mixture
have been pelletized under the pressure of 9 tons/cm$^{2}$ and heated
at 950$^{0}$C for 12 hours and at 1200$^{0}$C for 24 hours. 
The system has been finally sintered at 1400$^{0}$C for 24 hours with 
intermediate
grinding and pelletizing. Through out process the heating and cooling rate was 
maintained at 3$^{0}$C/minute and 2$^{0}$C/minute, respectively. 
Room temperature X-ray diffraction spectrum has beeen taken using Philips 
PW1710 diffractometer with Cu K$_{\alpha}$ radiation.
The XRD spectrum (Fig.1a) shows a well crystalline cubic spinel structure 
with lattice parameter (a) $\approx$ 8.4184 $\AA$.\\ 
The room temperature M\"{o}ssbauer spectrum (Fig.1b) in absence of magnetic 
field has been recorded in transmission geometry using a 25 mCi $^{57}$Co 
source in Rh matrix. 
The paramagnetic spectrum consists of two Lorentzian doublets arising from B 
site Fe$^{3+}$ ions and Fe$^{2+}$ ions respectively and a 
Lorentzian singlet due to A site Fe$^{3+}$ ions.
The fitted values of isomer shift (IS) and quadrupole 
splitting (QS) are +0.098 mm/sec, +0.339 mm/sec for B site Fe$^{3+}$ and
+0.808 mm/sec, 0.760 mm/sec for B site Fe$^{2+}$ ions respectively. The IS
value for the A site Fe$^{3+}$ ion is -0.181 mm/sec. The IS and QS
values are calculated with respect to Fe metal and with $\pm$0.002 mm/sec
error. The values are
in good agreement previously reported for ferrites \cite{Brand,Greenwood}.
The most probable cation distribution (with error =$\pm$ 0.001) obtained using
standard least square method is 
(Zn$^{2+}_{0.8}$Fe$^{3+}_{0.047}$Ti$^{4+}_{0.153}$)$_A$[Co$^{2+}_{0.2}$Fe$^{3+}_{1.530-{\delta}}$Fe$^{2+}_{0.023}$Ti$^{4+}_{0.247}$]$_B$O$_4$
(A: tetrahedral sites, B: octahedral sites). The term $\delta$ represent the
cations vacancy to maintain the charge neutrality in B site\cite{Millot} or if
$\delta$ is 0, then there is a possibility to obtain T$^{4+}$ and Ti$^{3+}$
mixtures in B sites \cite{Sugi}.

\subsection{Measurements}

The low field ac susceptibility in the temperature range 60K to 325K with ac
field $\sim$ 1 Oe and frequency range 37 Hz to 7.7 kHz and dc magnetization 
data under zero field cooled (ZFC) and field cooled (FC) condition have been 
recorded using home made magnetometer \cite{anindita}. 
For FC condition the cooling field and the
measurement field was maintained at the same value. 
The time dependence of remanent magnetization have been observed by 
field cooled condition with waiting time (t$_w$) $\approx$ 300 seconds. 
High field magnetic measurements have been performed using SQUID magnetometer.
The dc resistivity as a function of temperature in absence and presence of
magnetic field has been measured by two probe method 
using Kethley 6517A high resistance electrometer and Kethley 2001 multimeter
has been used for magnetoresistance measurement at T$\sim$ 240K to 315K. 

\section{Results}

\subsection{AC susceptibility}

The real ($\chi^\prime$) and imaginary ($\chi^{\prime\prime}$)
components of ac susceptibility (Fig.2) show the following features ,{\it
i.e.}, a small maximum at T$_{p}$ $\approx$ 240K, a plateau like behaviour in
the temperature range : 125 $< T >$ 240K, low temperature maximum at T$_f$ 
$\approx$ 125K, the decrease of ac susceptibility above T$_p$ and below T$_f$. 
These are the good characteristics for re-entrant magnetic phase transitions
in a magnetic system \cite{sanjoy}. 
The rounded maximum in $\chi^\prime$, $\chi^{\prime\prime}$ vs T at T$_p$
suggest that the magnetic clusters of different size, instead of individual 
spins, are showing magnetic response at T$_p$. 
Further, the absence of any significant shift of T$_p$ (Fig.
2) in the frequency range 37 Hz to 7.7 kHz of 1 Oe ac field suggests a strong 
ferromagnetic interactions inside the clusters \cite{Cohen}. If there was no
cluster size distribution, T$_p$ should have been represented the curie 
temperature T$_C$ of the system. However, keeping in mind the cluster size
distribution, it is expected that magnetic response of the largest clusters
will occur at T $> T_p$. In this case, it is not proper to call T$_p$ as the
Curie temperature T$_C$, where the system undergoes a paramagnetic to
ferromagnetic state on decreasing the temperature.
Therefore, the Curie temperature of the system is defined as the inflection 
point of $\chi^\prime$ at T$_C$ $\approx$ 270K and the difference between 
T$_C$ and T$_p$ ($\sim$ 30K) is due to the existence of various size cluster
below T$_C$ \cite{Liu}. 
If the magnetic clusters are assumed as the ferromagnetic domains, the absence 
of strong divergence in ac susceptibility data below T$_C$ suggest that 
the clusters in this system do not represent long range ordered 
ferromagnetic domains below T$_C$. However, the plateau like behaviour in
$\chi^\prime$, $\chi^{\prime\prime}$ vs T data below 240K suggest the
ferromagnetic ordering of the clusters \cite{sanjoy} with
the spin-spin correlation length (l) restricted to 
less than the size of the clusters due to various factors like spin canting,
random occupation of the magnetic and non-magnetic moments inside the
clusters. The plateau like behaviour may also be affected by the demagnetising 
field in the very thin semi-disc (length $>>$ breadth) shaped sample. However, 
the qualitative features of the sample will not signicantly changed due to the 
demagnetizing field effects. The effect of demagnetising field
becomes less important and is expected to be almost independent of sample 
shape in this system, as the cluster size is smaller than the percolation limit
where large number of spins are infinite long range ferromagnetic ordered
\cite{Sarkisan}.\\
Focussing on the low temperature maximum in $\chi^\prime$ and 
$\chi^{\prime\prime}$ (Fig. 2), it is observed that T$_f$ ($\approx$ 125K)
shows a significant frequency shift (marked by arrow).
This signals the appearance of more frustrated magnetic state with 
spin glass character at low temperature. 
The T$_f$(f) were estimated by first order derivative of $\chi^\prime$ with
rest to temperature and the data (with $\pm$ 2\% accuracy) were fitted with 
Vogel-Fulcher law
\begin{equation}
f = f_{0}exp^{-E_{a}/(T_{f}-T_{0})}
\end{equation}
The ln(f) vs 1/(T$_f$ - T$_0$) data are shown in Fig.2b inset.
The fitted parameters are f$_0$ $\approx$ 10$^{7}$ Hz, E$_a$ $\approx$ 220K and
T$_0$ $\approx$ 106K. These parameters suggest the freezing of clusters below
T$_f$. This justify to define T$_f$ is as cluster spin freezing temperature 
for this system. The spin glass feature at T $\leq$ T$_f$ suggest that short 
range antiferromagnetic interaction is becoming significant as T$\rightarrow$ 
T$_f$ from high temperature side and competing with ferromagnetic interactions
which is observed below 240K. The sharp decrease of $\chi^\prime$ and 
$\chi^{\prime\prime}$ suggest the increase of spin canting inside the clusters 
on lowering the temperature \cite{sanjoy,Sarkisan}. As a
consequence domain wall motion of the clusters is hindered below T$_f$ due to 
random anisotropy field \cite{Campbel}. 
\subsection{DC magnetization}

Fig.3a shows the zero field cooled (ZFC) and field cooled (FC) dc
susceptibility (M/H) with magnetic field H = 10 Oe to 85 Oe and 
temperature range T = 20K to 320K. 
If the temperature continues to decrease, the ZFC susceptibility   
shows a broad maximum 
at T$_{f}$ $\approx$ 125K. Below T$_f$, the ZFC susceptibility decreases upto 
our lowest measurement temperature 20K. The ZFC susceptibility is almost dc 
fields independent below 70K, as observed in other re-entrant system where low
temperature regime shows spin glass behaviour but strong field dependent 
ferromagnetic regime appear as the temperature increases \cite{Jonason}.
The inverse of zero field cooled susceptibility ($\chi_{dc}^{ZFC}$) vs T plot
above 275K shows an upward curvature and the data above 295K fit with
(Fig.3a inset) Curie-Weiss law 
\begin{equation}
\chi_{dc}= \frac{C}{T-\theta_{w}}
\end{equation}
The fitted parameters are Curie constant (C) $\approx$ 0.144 emu-K/g/Oe and
the asymptotic Curie temperature ($\theta_{w}$) $\approx$ + 270K. The intercept of the inverse ZFC susceptibility on the 
positive temperature axis suggest a dominant ferromagnetic interactions in
this system below T$_C$ $\approx$ 270K.\\
The FC susceptibility shows an weak magnetic irreversibility
at T$_{irr}$ ($< T_{C}$), which is followed by FC susceptibility maximum close
to T$_f$ with strong magnetic irreversibility at low temperature (see Fig.3a
inset for 10 Oe FC and ZFC data).
The decrease of FC susceptibility below T$_f$ can be attributed due to the
strong antiferromagnetic interactions or due to the local random anisotropy 
below below T$_f$ \cite{sanjoy}. Fig.3b inset shows that irreversibility
temperature T$_{irr}$ and cluster spin freezing temperature T$_f$ decreases 
with increasing magnetic field. The appearance of an weak magnetic 
irreversibility below T$_C$ and strong magnetic irreversibility below T$_f$
suggest the re-entrant magnetic behaviour \cite{sanjoy} in this system. 

\subsection{Magnetic hysteresis}

We have shown magnetic field (H) dependence of ZFC isothermal magnetization
(M) in Fig.4. 
The system shows some remarkable feature in M vs H plot. The 10K data show
an S shape at low field regime, while the 30K and 60K data 
show a field induced magnetic transition at H$_{cf}$ $\sim$ 
1 Tesla and 1.5 Tesla, respectively. It is also
observed that after 4 quadrant field cycling (0 $\rightleftharpoons$ H$_{max}$
and 0 $\rightleftharpoons$ H$_{min}$), the field induced
transition occurs at higher fields, {\it i.e.}, H$_{cf}$ $\sim$ 1.7 Tesla and 
3 Tesla 
for 30K (Fig.4b) and 60K, respectively. The 100K do not show any field induced
transition at + H axis but small field induced transition is observed at
$\approx$ -1 Tesla. At T $\geq$ 150K no field induced transitions are observed.
Similar phase induced transitions have been observed in a variety of
disordered magnetic materials where the materials segregates into two
distinguishable electronic states that coexist within the
same crystallographic phase \cite{Mahen,sanjoy,Gordon,Nojri,Filpov,Nigam}.
The most reasonable explanation for such type of field induced transition is
that the system undergoes an antiferromagnetic to metastable ferromagnetic
state at H $> H_{cf}$. This is further confirmed that there is no long range
ferromagnetic order in the system, rather the antiferromagnetic interactions
promotes strong spin canting inside the clusters below 100K and the
ferromagnetic state above T$_f$ ($\approx$ 125K) is very similar to canted 
ferromagnets \cite{sanjoy,Nigam}. 
The same features are observed in the M vs H plots where
magnetization shows lacks of saturation even upto 8 Tesla.
The low value of coercive
field above 125K and no magnetic hysteresis above 150K suggest ferromagnetic
regime of the sample. However, the rapid increase of coercive field below 100K 
(Fig.5 inset right scale) definitely suggest the blocking of domain wall
motion in the spin canting state \cite{sanjoy,Campbel}.\\
By inspecting the positive slope of H/M vs M$^{2}$ at T $\leq$ 250K, we
assumed that the paramagnetic to ferromagnetic phase transition in this system
is second order \cite{Filpov,Mira}.
We have calculated the spontaneous magnetization (M$_S$) using modified Arrot 
plot (Fig.5)
\begin{equation}
M^{1/{\beta}}(H) \propto (H/M)^{1/{\gamma}}
\end{equation}
A self consistent method to obtain the best fitted parameters was considered
\cite{Mira}.
In this procedure, first we have estimated the values of critical parameters
using 250K data and modified Arrot plots were constructed using this
parameters for all the temperatures.
The linear extrapolation of high field magnetization data to M$^{1/{\beta}}$
axis gives the M$_S$ value and the linear extrapolation to
(H/M)$^{1/{\gamma}}$ axis gives the inverse of initial susceptibility. The
M$_S$ values were applied to the equation: M$_S$ (T) $\propto$ (T$_C$ -
T)$^{1/{\beta}}$ for T $< T_{C}$. The modified Arrot plots were reconstructed
using the obtained values of $\beta$. Finally, we obtained the best fitted
values as : $\beta$ = 1.03$\pm$0.02, $\gamma$ = 0.70$\pm$0.01 and T$_C$
$\approx$ 265K $\pm$2K. 
The exponent values are close to that reported values for Heisenberg 
ferromagnets with strong magnetic disorder \cite{Belachi}.
The temperature
dependence of the M$_S$ (Fig.5 inset left scale) suggest that M$_S$ value at
10K ($\sim$ 46 emu/g) is
less than the values obtained at 30K and 60K ($\sim$ 53 emu/g). This is due 
to the spin canting effect at low temperature.

\subsection{Time and temperature dependence of dc magnetization}

We have investigated the time dependence of field cooled remanent
magnetization for better understanding of competition between equilibrium and 
non-equilibrium dynamics in the system. The experimental data (point symbol)
in Fig.6a
show a clear and systematic change of time dependent curvature from concave 
down (at T $\leq$ 130K) to concave up 
as the measurement temperature increases. We have found that 
the remanent magnetization decay at low temperature regime (T $\leq$ 130K) is
best fitted by the superposition of a pure stretched exponential and a
 constant term as
 \begin{equation}
 M_{R}(t) = M_{0} + M_{1}exp^{-(t/\tau)^{n}}
 \end{equation}
where M$_0$, M$_1$, $\tau$ and n are fitted parameters (shown in Table 1). 
The M$_0$ parameter represents intrinsic ferromagnetic contribution and M$_1$
relates to a glassy component at the measurement temperature \cite{Cohen}.
Above, 130K the remanent magnetization decay is not well fitted by equation
(3). However, the best fit of 142K data require the product of a power law
term with the stretched exponential function as 
 \begin{equation}
 M_{R}(t) = M_{0} + M_{1}t^{-\alpha}exp^{-(t/\tau)^{n}}
 \end{equation}
where M$_0$, M$_1$, $\alpha$, $\tau$ and n are the fitted parameters (shown in
Table 1). In general, the power law decay of remanent
magnetization has been considered as a weak time dependent function
and represents an equilibrium spin dynamics of a ferromagnetic
 system \cite{Roso}, where as the stretched exponential function represent the 
non-equilibrium slow spin glass dynamics \cite{Jonason}. 
The relaxation of the system above the cluster spin freezing temperature T$_f$
$\approx$ 125K suggest that it is not in a true ferromagnetic state. The
disorder and frustration in the ferromagnetic state takes
into account the two competitive times scale simultaneously, one is due to 
intra-cluster dynamics \cite{Cohen} and second one is due to the cluster
growth in presence of the frustration \cite{Nam}. 
According to a model proposed by Chamberlin and Haines \cite{Chamber} that 
relaxation rate ($\omega_{s}$) of a cluster with size 's' is related as 
\begin{equation}
\omega_{s} \sim exp(C/s)
\end{equation}
where C is a constant.  
This equation suggest that as the cluster size decreases, the relaxation rate
will be faster. This is possible as the thermal activated process, on
increasing temperature, will decrease the cluster size. 
In the temperature range T$\geq$ 166K to 251K, the time dependence of
remanent magnetization follows a power law, similar to that has been observed
in a re-entrant ferromagnet \cite{Roso} by the equation
\begin{equation}
 M_{R}(t) = M_{0} + M_{1}t^{-\alpha}
 \end{equation}
with the fitted parameters shown in Table 1.
It has been found that the remanent 
magnetization at 280K is practically independent of time 
(M$_{R}$(t) $\approx$ M$_0$) and the small value of M$_0$ is due to the short
range interacting clusters in the paramagnetic state \cite{Belachi}.\\
The temperature dependence of the field cooled remanent magnetization 
(M$_{TRM}$), after removing the cooling field $\approx$ 60 Oe, is shown 
in Fig.6b. The data clearly show a change in the
nature of remanent magnetization decay at a tempearure T$_S$ $\approx$ 130K.
We have found that TRM data at T$\leq$ 130K decay as
\begin{equation}
M_{TRM} \propto exp(-{\alpha}T)
\end{equation}
where as the TRM data at T $>130K$ follow a simple power law decay as
\begin{equation}
M_{TRM} \propto T^{-{\beta}}
\end{equation}
The implication of the above
two equations is that the thermal activated process slowly reduces the glassy
behaviour exhibits below T$_S$ in comparison with the fast decay in the ferromagnetic
state. However, the non-zero value of M$_{TRM}$ at 280K is
consistent with our assumption that it is due to the short range interacting
clusters in the paramagnetic state.
\subsection{Temperature dependence of dc resistivity}

As a complimentary support for the existence of re-entrant magnetic state, we 
have measured the resistivity ($\rho$) as a function of temperature and 
magnetic field. Fig.7a shows $\rho$ vs T plot at zero and 7.8
kOe magnetic field. 
The zero field resistivity rapidly increases below 270K and shows slow increase
below 200K that indicates the presence of ferromagnetic ordering below 200K. 
The high resistivity values even below 200K suggest that the ferromagnetic 
interaction is weak in this system \cite{Sundar}. The strong increase of 
resistivity below 100K suggest that antiferromagnetic interaction is
dominating over the ferromagnetic one due to strong spin canting effect at
low temperature.
The resistivity follows an exponential function as $\rho$ $\sim$
exp$^{E_{g}/kT}$ in both the regimes, suggests semi-conducting behaviour of
the system. The fitted value of the activation 
energy (E$_g$) is decreasing from $\approx$ 0.239 eV (for T $> 200K$) to 
5.98 meV (for T $<$ 200K). This informs that the paramagnetic like disordered 
state become more ordered (ferromagnetic) below 200K \cite{Ahmed}.\\ 
It is observed (Fig.7a) that as the temperature decreases below 320K,
resistivity under field ($\rho$ (7.8 kOe)) remains greater than $\rho$ (0) 
value upto T$\approx$ 270K. Then
$\rho$ (7.8 kOe) $\sim$ $\rho$ (0) upto 240K but increases upto T$_{SM}$ 
$\approx$ 175K, where the system shows field induced semiconductor to metal 
transition. On further decreasing temperature,
$\rho$ (7.8 kOe) decreases before showing a sharp increase below  
the field induced metal to semiconductor transition at T$_{MS}$ $\approx$ 80K.
This is a re-entrant semi-conducting (S-M-S) behaviour \cite{Good} in our 
system. The magnetoresistance (MR) calculated (from Fig.7a) using the formula
${\Delta}{\rho}$/$\rho_{0}$[=($\rho_{7.8 kOe}$-$\rho_{0}$)/$\rho_{0}$] 
shows $\approx$ 80\% (-ve) change at T$_{MS}$, $\approx$ 20\% (-ve) at
T$_{SM}$, $\approx$ 40\% (-ve) at 200K and positive MR above 270K.\\
The comparison of resistivity ($\rho$ (H)) with the ac susceptibility data 
(Fig.7a right scale) shows that the positive
magnetoresistance occurs in the paramagnetic regime (T$>T_{C} {\approx}$
270K), the negative MR is observed below T$_p$ where the clusters show
ferromagnetic ordering and the magnetoresistance again decreases in the spin
glass regime (T $<$ 100K).
The difference between the ac susceptibility maximum at T$_p$ $\approx$ 240K 
and the $\rho$ (H) maximum at 175K probably due to the existence of different 
size clusters in the system. Depending on the size and random distribution of
magnetic and non-magnetic ions inside the clusters, the proportion of  
ferromagnetic (metallic) and paramagnetic (semi-conducting) contributions will 
vary from cluster to cluster in the system. 
These clusters are called magnetic polarons or ferrons \cite{Good,Mahen}.
Some of the clusters, which are paramagnetic in nature, will show
semi-conducting and most of the clusters, which are ferromagnetic in nature,
will show metallic behaviour. As a result, instead of showing $\rho$ (H)
maximum near to T$_p$ or T$_C$, the system shows $\rho$ (H) maximum at 
$\approx$ 175K. This behavior is in contrast to the conventional CMR
perovskites where $\rho$ (H) occurs near to T$_C$ \cite{Mathur}. The
resistivity in presence of magnetic field, therefore, confirms that there is
no long range ferromagnetic order in this system and the spin-spin correlation
length varies from cluster to cluster.
Below 175K, the high magnetic field strongly reduces the paramagnetic 
effect and increases the size of ferromagnetic clusters. 
As the spin-spin correlation length increases in presence of
field, the scattering of the charge carrier decreases and the enhancement 
of the electron hopping inside the clusters show metallic behaviour in the 
system. The appearance of canted spin
glass state at T $< T_{f}$, where antiferromagnetic interaction dominates,
will the spin-spin correlation length and increases the electron scattering 
process inside the clusters. The localization of electrons
in spin canted states show sharp increase of resistivity \cite{Mahen}.\\
We have also
measured the magnetoresistance (MR), after zero field cooling from 300K to the
measurement temperature ($\sim$ 240K to 315K). The temperature has been kept 
constant with
maximum fluctuation $\pm$ 0.1K during the measurement period. 
We find (Fig.8) an appreciable change in magnetoresistance
${\Delta}{\rho}$/$\rho_{0}$[=($\rho_{H}$-$\rho_{0}$)/$\rho_{0}$] for magnetic
field $\pm$ 7.8kG. It is to be noted (Fig.8a) that the MR is negative at 247K
and suggests dominant ferromagnetic contribution. For T $\geq$ 252K, the MR 
initially shows positive value and then crossover to negative value as the 
applied field increases. This indicates a competition between ferromagnetic
and paramagnetic response of the clusters below T$_C$. On further
increasing the measurement temperature, the system shows 
positive MR (Fig.8b) at T $\geq$ T$_C$ ($\approx$ 270K) (Fig.8c) in the 
paramagnetic regime. The positive MR in the paramagnetic state may be due to 
the Lorentz force \cite{Belov}, which causes bending ('twisting') of the 
conduction electron's path, or grain boundary scattering as expected for 
polycrystalline sample in the paramagnetic state \cite{Mathur}.  
To confirm about the cross over from negative to positive MR near to Curie
temperature (T$_C$), we have continued the MR measurement at a particular 
temperature for more than one field cycling( 0 Oe $\rightarrow$ 7.8kOe $\rightarrow$ 0 Oe $\rightarrow$ -7.8 kOe
$\rightarrow$ 0 Oe). Fig.9a shows that the MR
at T = 252K ($<T_{C}$) continuously decreases on 
cycling the field for three
times, in contrast to the continuously increasing trend (Fig.9c) at 289K 
($>T_{C}$). However, if we look at the 272K (near to T$_C$) data (Fig.9b),
the MR value never gets back to the starting point after the one
cycle field application. The continuation of increasing or decreasing trend of 
magnetoresistance is related to relaxation \cite{Rivas} or memory effect 
\cite{Mathur,Gordon} in the sample.
\section{Discussion}

The magnetic and complementary resistivity measurements of
Co$_{0.2}$Zn$_{0.8}$Fe$_{1.6}$Ti$_{0.4}$O$_4$ spinel oxide give the signatures 
of re-entrant spin glass behaviour, eventhough, the existence of a true
re-entrant spin glass phase is still a matter of debate for 3-dimensional 
system \cite{Jonason,Sato}. The experimental results will be discussed by
assuming the existence of various size clusters. The ferromagnetic behaviour
has been understood by treating these clusters as magnetic domains, inside of
which all the spins are not necessarily be infinite long range ordered due to
spin canting effects, random distribution of magnetic and non-magnetic ions 
and different proportion of ferromagnetic and antiferromagnetic
interactions inside the clusters. 
The temperature dependence of ac susceptibility data suggest that domain
formation occurs above 240K. As the temperature decreases, canted spin 
structure is favoured inside the domains due to antiferromagnetic interactions.
As a result, the random anisotropy field
introduced by Dzyalosinsky-Moriya type interactions $\sim$ 
$\vec{S_i}$x$\vec{S_j}$ will hinder the (cluster) domain wall movement and 
decrease the low field ac susceptibility at low temperature \cite{sanjoy}. 
The frequency dependence of ac susceptibility ($\chi^\prime$, 
$\chi^{\prime\prime}$), the weak irreversibility (between FC and ZFC 
susceptibility) below T$_C$ ($\approx$ 270K) and strong irreversibility below 
T$_f$ ($\approx$ 125K) give the characteristic feature of typical re-entrant 
ferromagnet \cite{Roso}. The re-entrant character of the system is further
confirmed by the time dependence of the field cooled remanent magnetization
experiments. The slow spin dynamics at T$\leq$ 130K
suggest the non-equilibrium canted  spin 
glass state, where as an equilibrium spin dynamics is dominating over the 
non-equilibrium dynamics in the ferromagnetic phase (weak time dependent) 
and the spin dynamics, then, approaches toward a time independent disordered 
(paramagnetic) phase as observed at 280K. The relaxation data along with field
dependence of magnetization data also confirm that there is no long range
ferromagnetic order, rather the system is a canted ferromagnet.\\ 
The most attractive feature of the present system is that the
magnetic ordering and transport properties are highly correlated. 
The M\"{o}ssbauer analysis suggests the existence of Fe$^{3+}$ and Fe$^{2+}$
ions and charge compensating defects (due to valence mismatch between 
Ti$^{4+}$ and Fe$^{3+}$ ions) like cation vacancies in B site \cite{Millot}.
The charge compensating defects can give rise ferromagnetic interactions by 
renormalizing the concentration of frustrated bonds \cite{Aha1}.
Simultaneously, the charge fluctuations among the cations (Ti$^{4+}$/Fe$^{2+}$ 
to Ti$^{3+}$/Fe$^{3+}$) in B site clusters will enhance the electron (polaron) 
hoping mechanism \cite{Sugi} and consequently the system shows large
electrical conductivity. On the otherhand, tetravalent non-magnetic 
Ti$^{4+}$ ions substitution in place of magnetic Fe$^{3+}$ ions makes the
inter-sublattice interactions Fe$_{B}^{3+}$-O$^{2-}$-Fe$_{A}^{3+}$ almost
negligible and the magnetic interactions depends on the B site superexchange 
interactions like Fe$_{B}^{3+}$-O$^{2-}$-Fe$_{B}^{2+}$,
Fe$_{B}^{3+}$-O$^{2-}$-Ti$_{B}^{3+}$,
Fe$_{B}^{2+}$-O$^{2-}$-Ti$_{B}^{4+}$ etc. From the various measurements, it is
observed that although there exists both ferromagnetic and antiferromagnetic
superexchange interactions, the dominant one is the ferromagnetic
interactions which is observed from the intercept of the inverse of 
susceptibility on positive temperature axis. The appearance of metallic like
resistivity behavour in presence of dc field at 80K to 175K range strongly 
suggest that magnetic exchange interactions are also affecting the electrical 
properties of the sample.

\section{Conclusions}

In conclusion, the present system 
Co$_{0.2}$Zn$_{0.8}$Fe$_{1.6}$Ti$_{0.4}$O$_4$ exhibits re-entrant magnetic
phase transitions, {\it i.e.} paramagnetic to ferromagnetic state at 
T$_C$ $\approx$ 270K and ferromagnetic to canted spin glass state at T$_f$ 
$\approx$ 125K. The re-entrant behaviour in magnetic and in electrical
resistivity data is attributed due to the cation vacancy and charge 
fluctuation effects inside the ferromagnetic clusters (magnetic polarons). 
The ferromagnetic state is not a typical long range ordered
type, rather than a canted ferromagnetic one, where longitudinal spin
components show ferromagnetic order and transverse spin components 
show spin glass order at low temperature. Without Neutron diffraction 
experiment, it is very difficult to say whether the ferromagnetic order of 
longitudinal spin components is still maintaining in canted spin glass state. 
However, the variation of exchange interactions,
random distributions of cations from cluster
to cluster may give rise inhomogeneity in electronic phase, {\it i.e.} a
mixture of metallic (ferromagnetic) and semi-conducting (antiferromagnetic)
components within the same cluster, eventhough, the
system is in same crystallographic (cubic spinel) phase.
\vspace{0.6 truecm}\\
\noindent Acknowledgement:
One of the authors RNB thanks to The Council of Scientific and Industrial
Research (CSIR, New Delhi, India) for providing fellowship [F. No.
9/489(30)/98-EMR-I]. We also thank C. Bansal, S. sarkar and S. Kumar for
helping in M\"{o}ssbauer data support.

\newpage
Table 1: The fitted parameters M$_0$, M$_1$, n, $\tau$ and $\alpha$ of
equations 4, 5 and 7 while fitted with time dependence of remanent
magnetization data. For detais see text.\\
\begin{tabular}{c c c c c c}
\hline\hline
T(K) & M$_0$ $\pm{\Delta}$ & M$_1$ $\pm{\Delta}$ & n $\pm{\Delta}$ & $\tau$ &
$\alpha$$\pm{\Delta}$ \\\hline
80K & 0.164$\pm$0.002 & 1.327$\pm$0.001 & 0.068$\pm$0.001 & 10$^5$ s & -\\
90K & 0.260$\pm$0.001 & 0.728$\pm$0.001 & 0.227$\pm$0.001 & 10$^5$ s & -\\
109K & 0.270$\pm$0.004 & 0.679$\pm$0.001 & 0.158$\pm$0.001 & 10$^5$ s & -\\
120K & 0.373$\pm$0.003 & 0.390$\pm$0.002 & 0.219$\pm$0.001 & 10$^5$ s & -\\
130K & 0.259$\pm$0.005 & 0.573$\pm$0.001 & 0.130$\pm$0.001 & 10$^5$ s & -\\
142K & 0.272$\pm$0.002 & 0.511$\pm$0.001 & 0.090$\pm$0.001 & 10$^5$ s &
0.006$\pm$0.0005\\
166K & 0.463$\pm$0.003 & 0.228$\pm$0.001 & - & - & 0.115$\pm$0.0001\\
189K & 0.283$\pm$0.003 & 0.069$\pm$0.001 & - & - & 0.276$\pm$0.0001\\
216K & 0.150$\pm$0.002 & 0.063$\pm$0.001 & -& - & 0.118$\pm$0.0002\\
251K & 0.109$\pm$0.001 & 0.015$\pm$0.001 & - & - & 0.194$\pm$0.0001\\
280K & - & - & - & -& -\\
\hline\hline
\end{tabular}
\newpage
\centerline{Figure Caption}
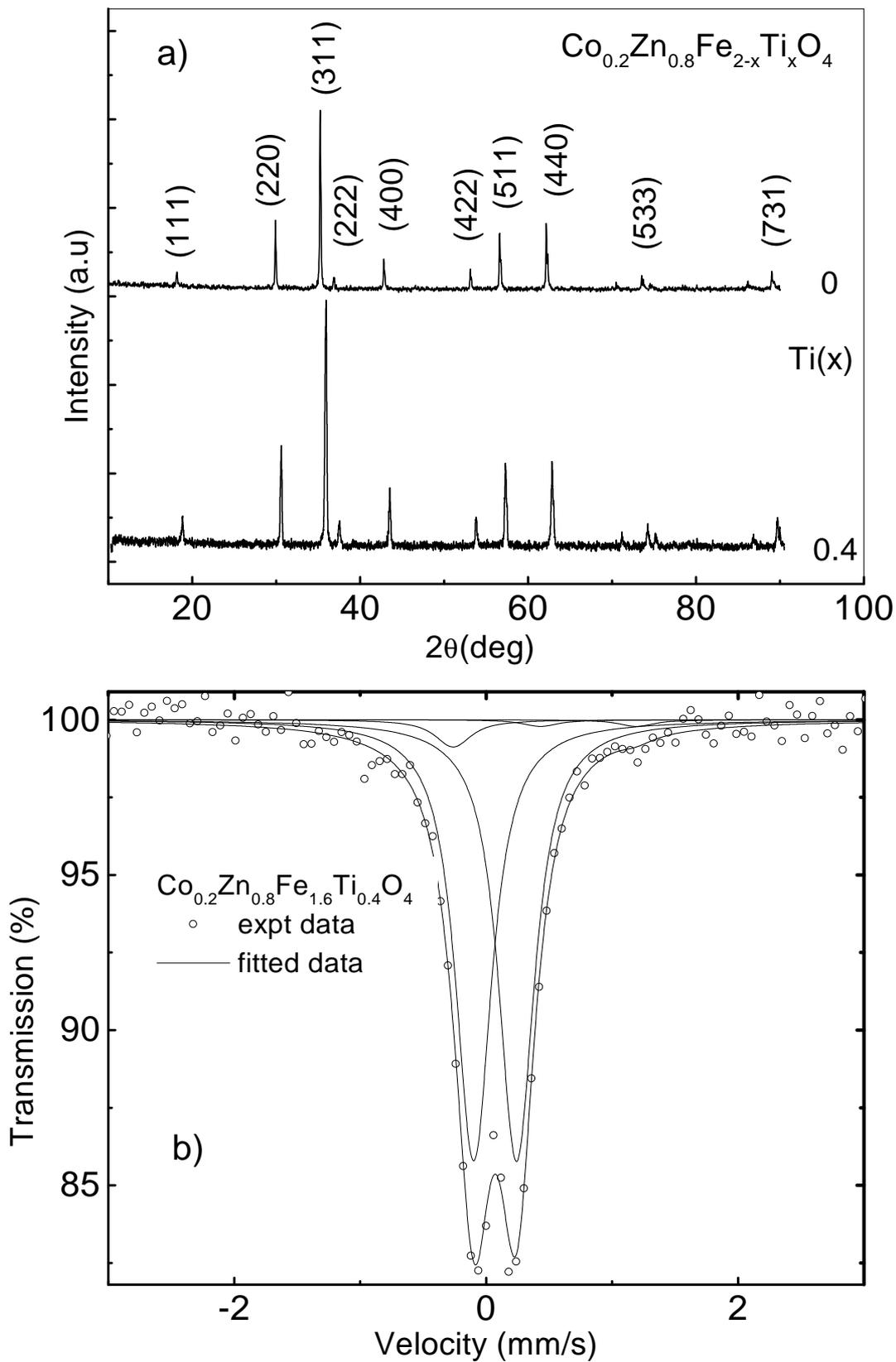
Fig.1a XRD data for Co$_{0.2}$Zn$_{0.8}$Fe$_{2-x}$Ti$_x$O$_4$ spinel oxide, b)
M\"{o}ssbauer spectrum recorded in absence of magnetic field at 300K. Solid
point: experimental data, dotted/solid line is the fitted data using least 
square method\\
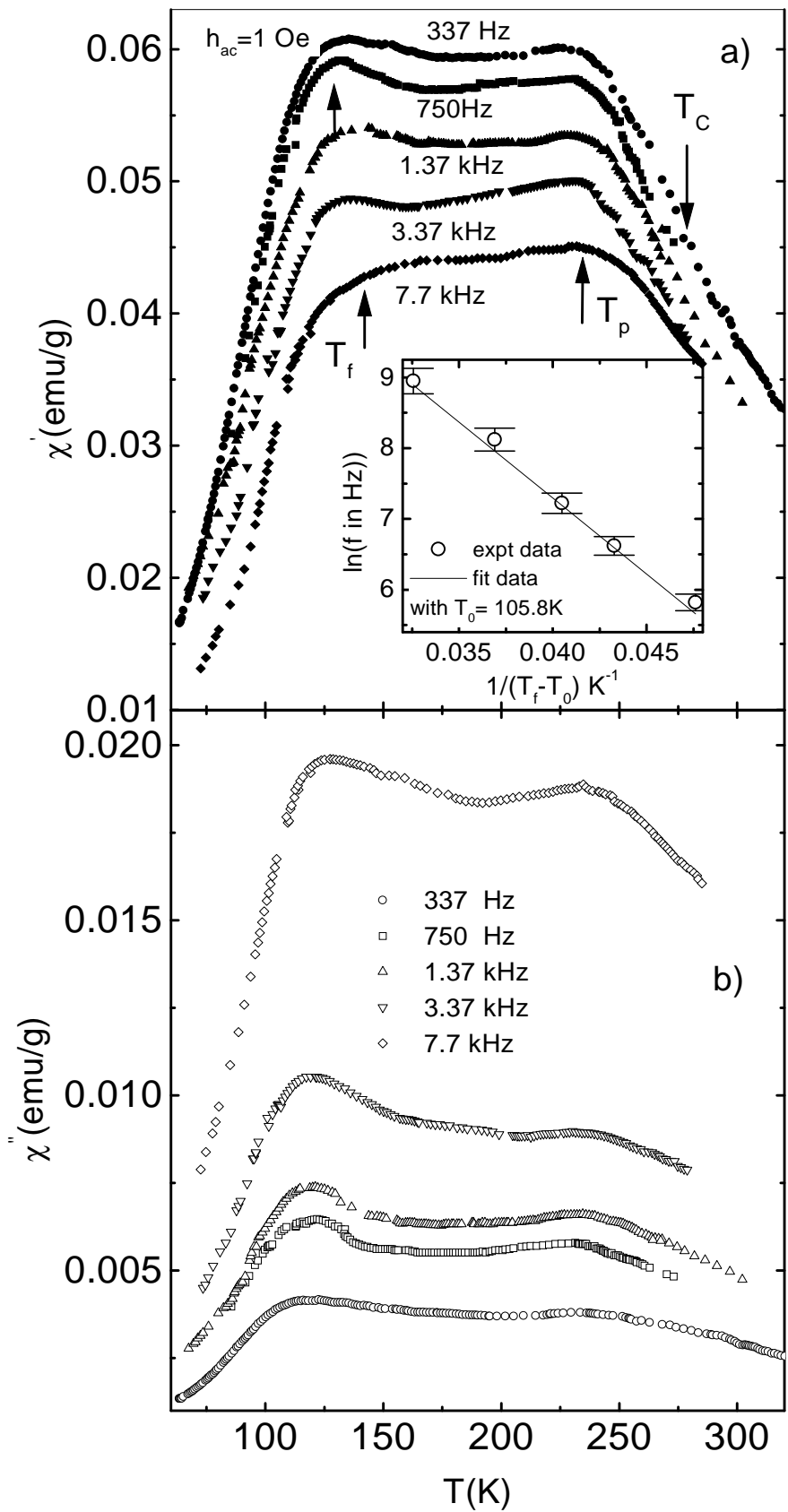
Fig.2 Temperature dependence of real ($\chi^\prime$) and imaginary
($\chi^{\prime\prime}$) component of ac susceptibility measured at 1 Oe ac
field in the frequency range 337 Hz to 7.7 kHz. The inset Fig. shows the expt
data fitted with Vogel-Fulcher law\\
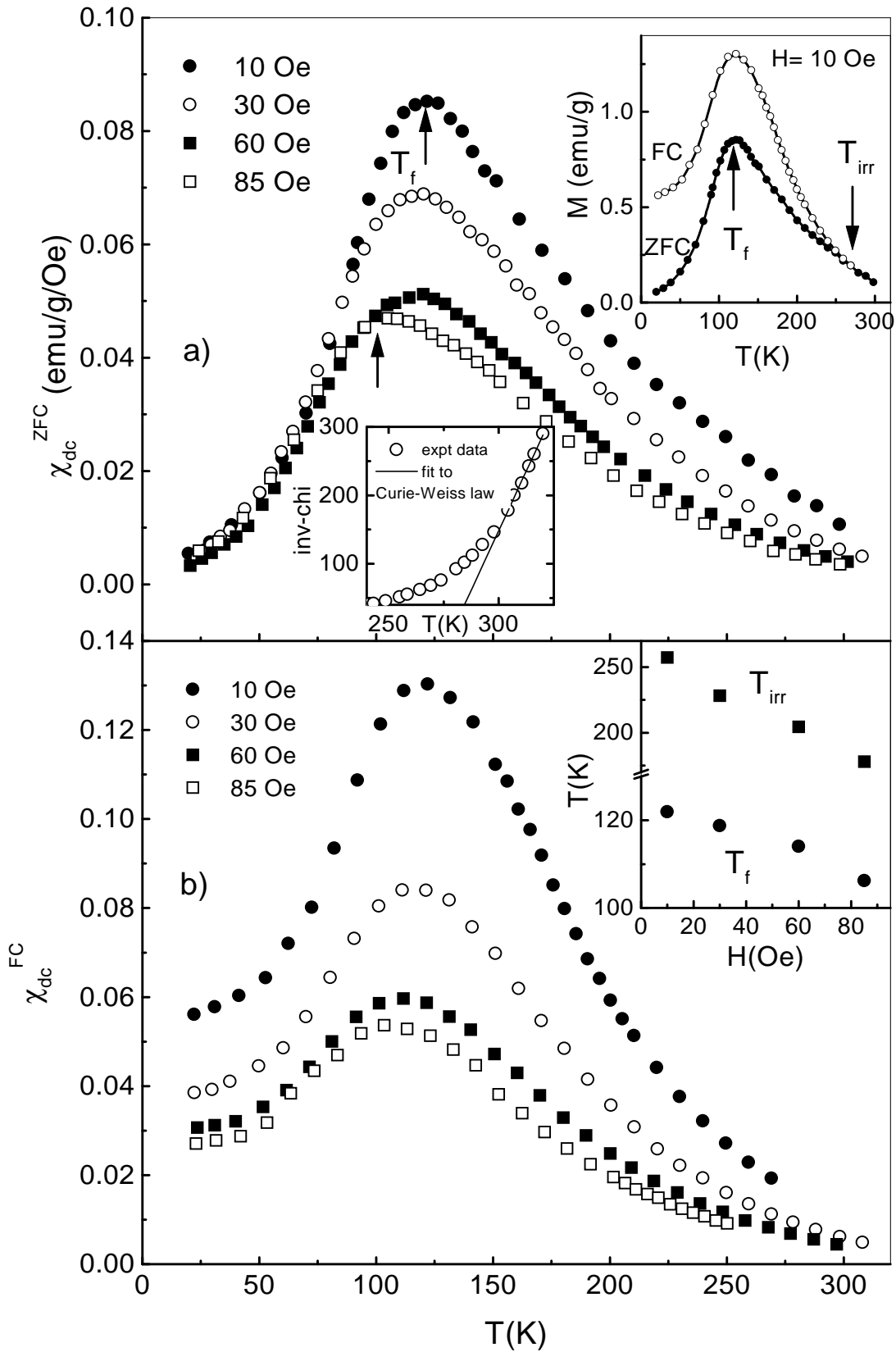
Fig.3 a) Zero field cooled dc susceptibility vs temperature measured at
different fields. The inset shows the zero field cooled (ZEC) and field cooled
(FC) magnetization vs temperature measured (top corner) and Curie-weiss law
fit to expt data (at close interval) measured at 10 Oe. b) field cooled dc
susceptibility vs temperature and the cluster
freezing temperature (T$_f$) and irreversible temperature (T$_{irr}$) at
different fields.\\
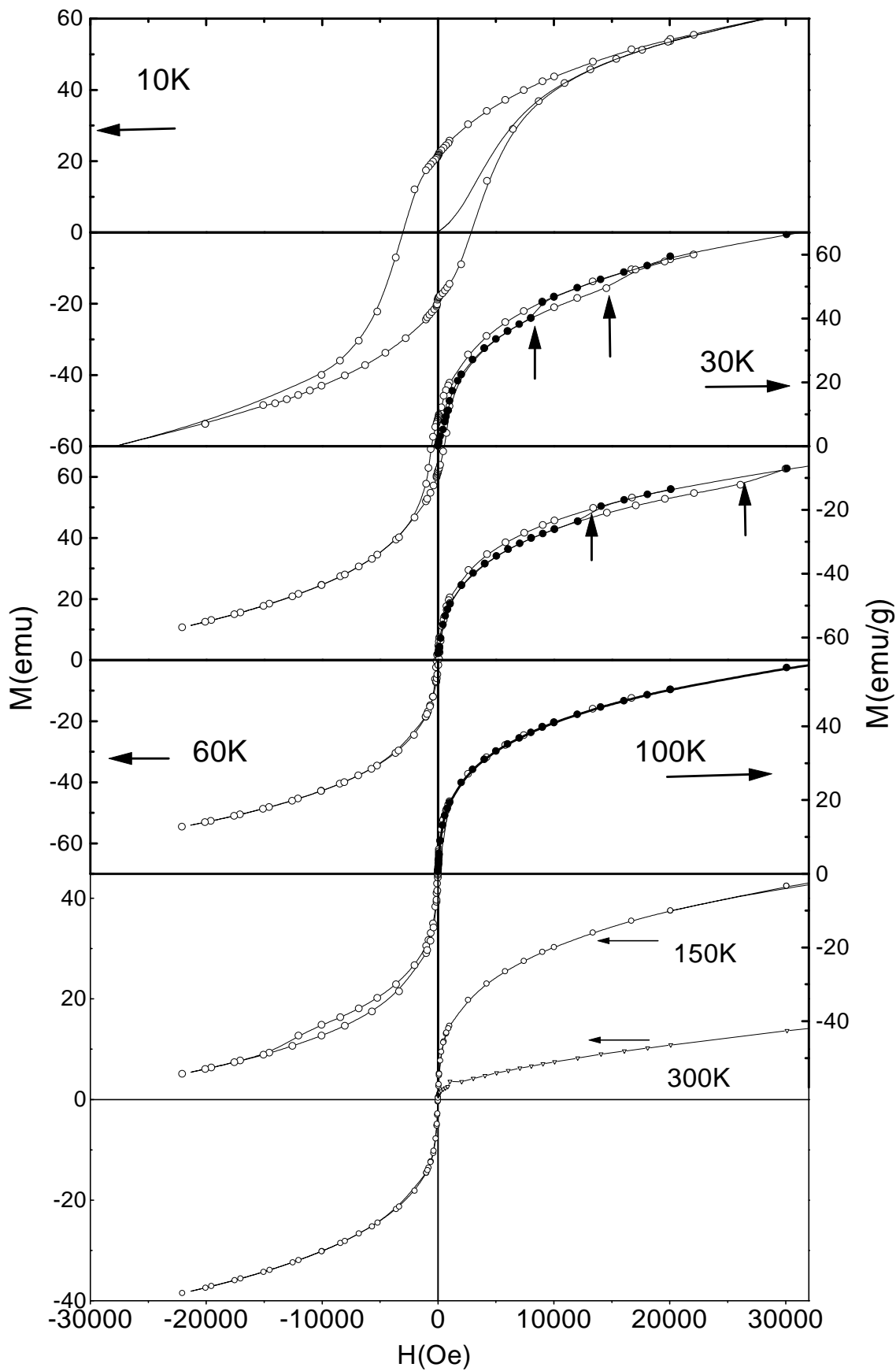
Fig.4 Hysteresis loop shown for 10K to 300K. The solid line represents the
increase of M as H increases 0 to 7 tesla. The open symbol shows the loop. The
arrow up indicate the field where field induced transition ocurs. The
horizontal arrow represent the M (axis) values at the temperature indicated.
For details see in the text.\\
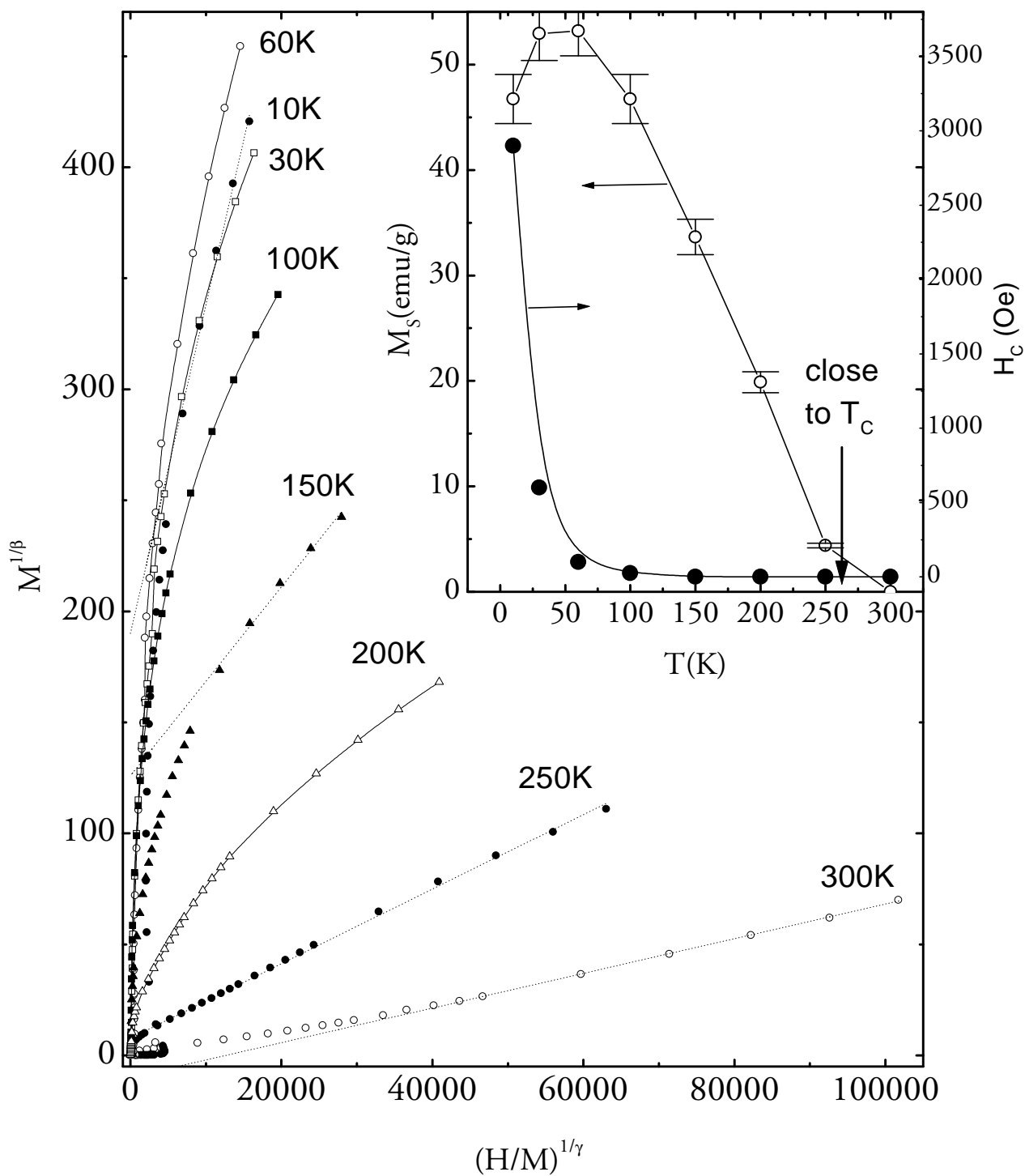
Fig.5 Modified Arrot plot (M$^{1/\beta}$ vs (H/M)$^{1/\gamma}$ with $\beta$ =
1.03 and $\gamma$ 0.7) for temperature 10K to 300K. The dotted line is the
linear extrapolation of M$^{1/\beta}$ (for H$>$ 3 Tesla) to H = 0 axis to
calculate the spontaneous magnetization and the solid lines are drawn for
guide to eye. Inset shows temperature dependence of spontaneous magnetization
(M$_S$) and coercive field (H$_C$)(right scale).\\
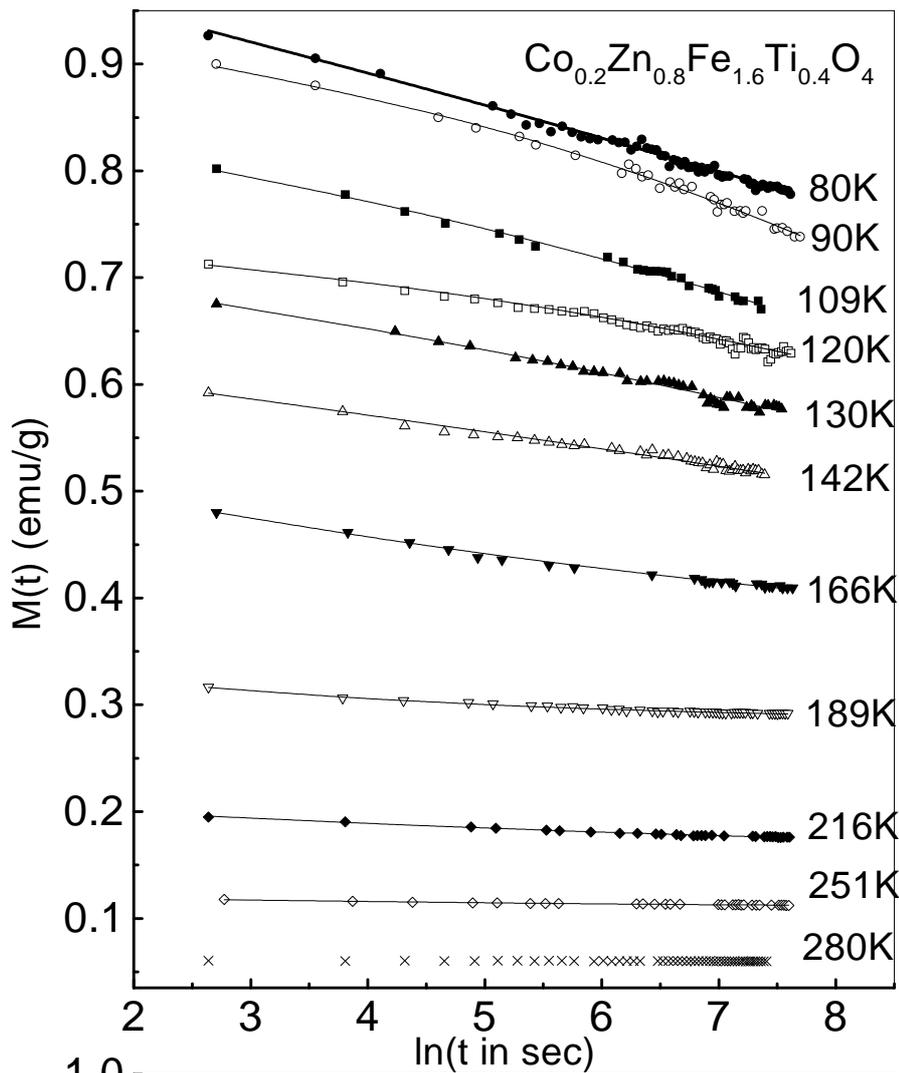
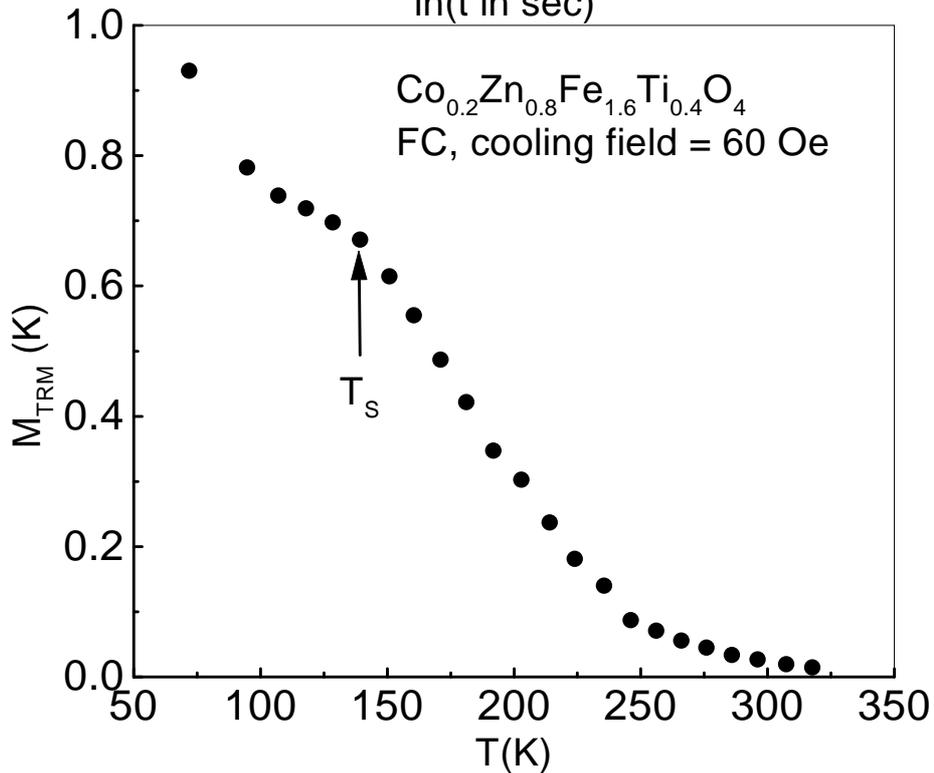
Fig.6 a) Time dependence of remanent magnetization measured at different
temperatures, b) Temperature dependence of remanent magnetization. T$_S$ is
the temperature which separate two decay regions.\\
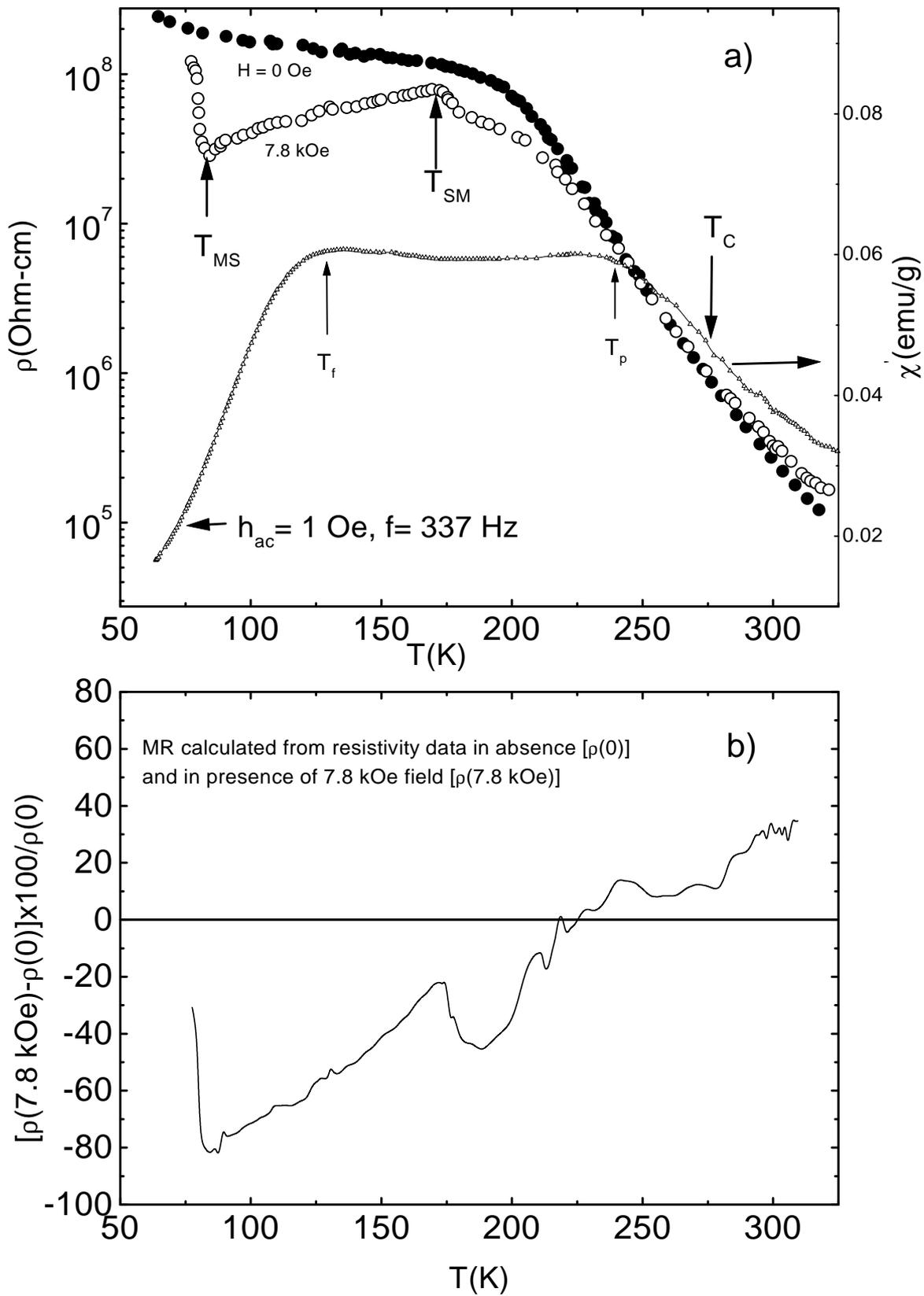
Fig.7 a) Temperature dependence of resistivity at 0 Oe and 7.8 kOe and ac
susceptibility at 1 Oe, 337 Hz (right scale).
b) magnetoresistance calculated by subtracting 7.8 kOe data from 0 Oe
data. For T$_f$, T$_p$, T$_{MS}$ and T$_{SM}$ see in the text.\\
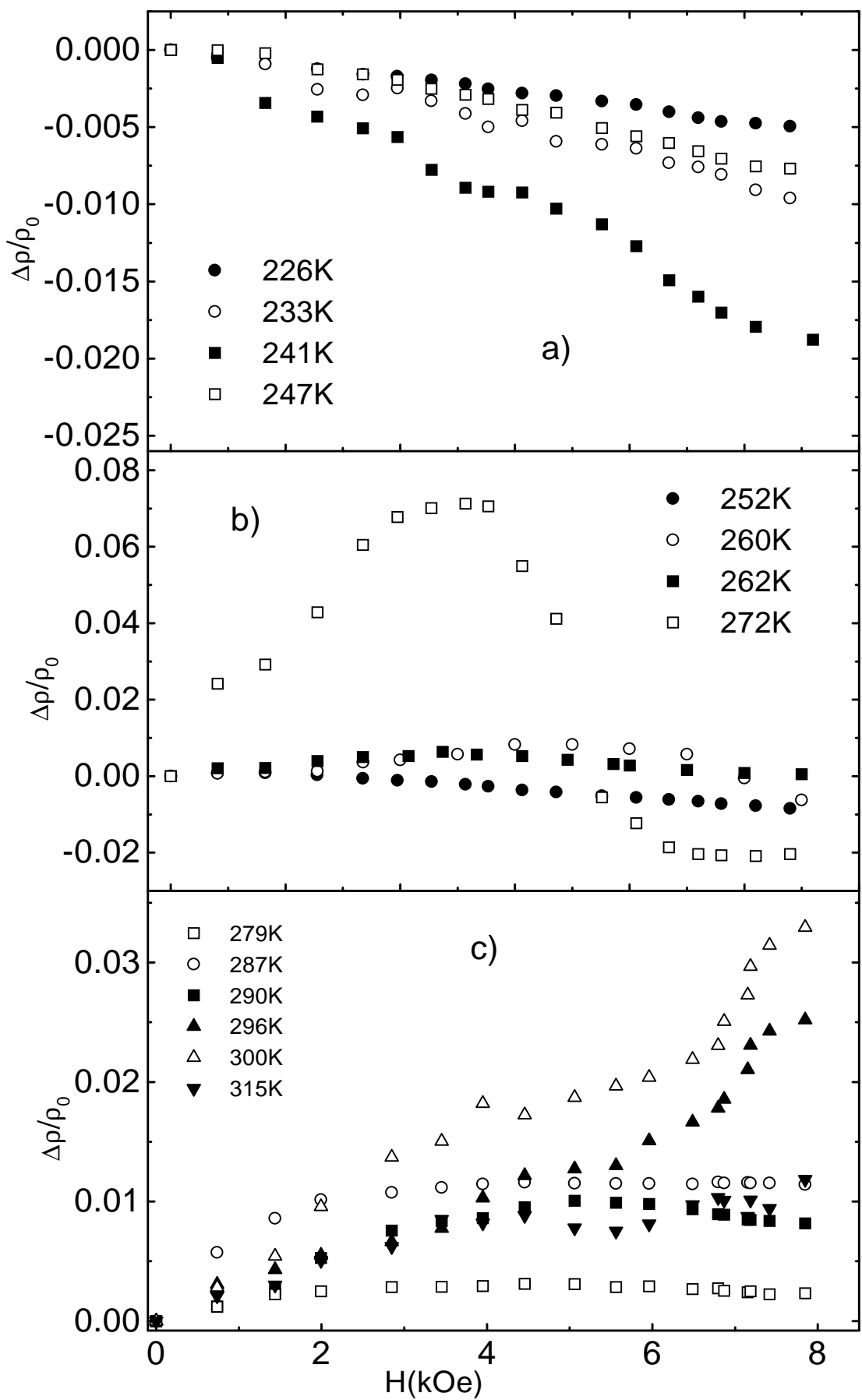
Fig.8 Magnetoresistance measured at different temperatures with maximum field
8 kOe for Co$_{0.2}$Zn$_{0.8}$Fe$_{2-x}$Ti$_x$O$_4$ spinel
oxide.\\
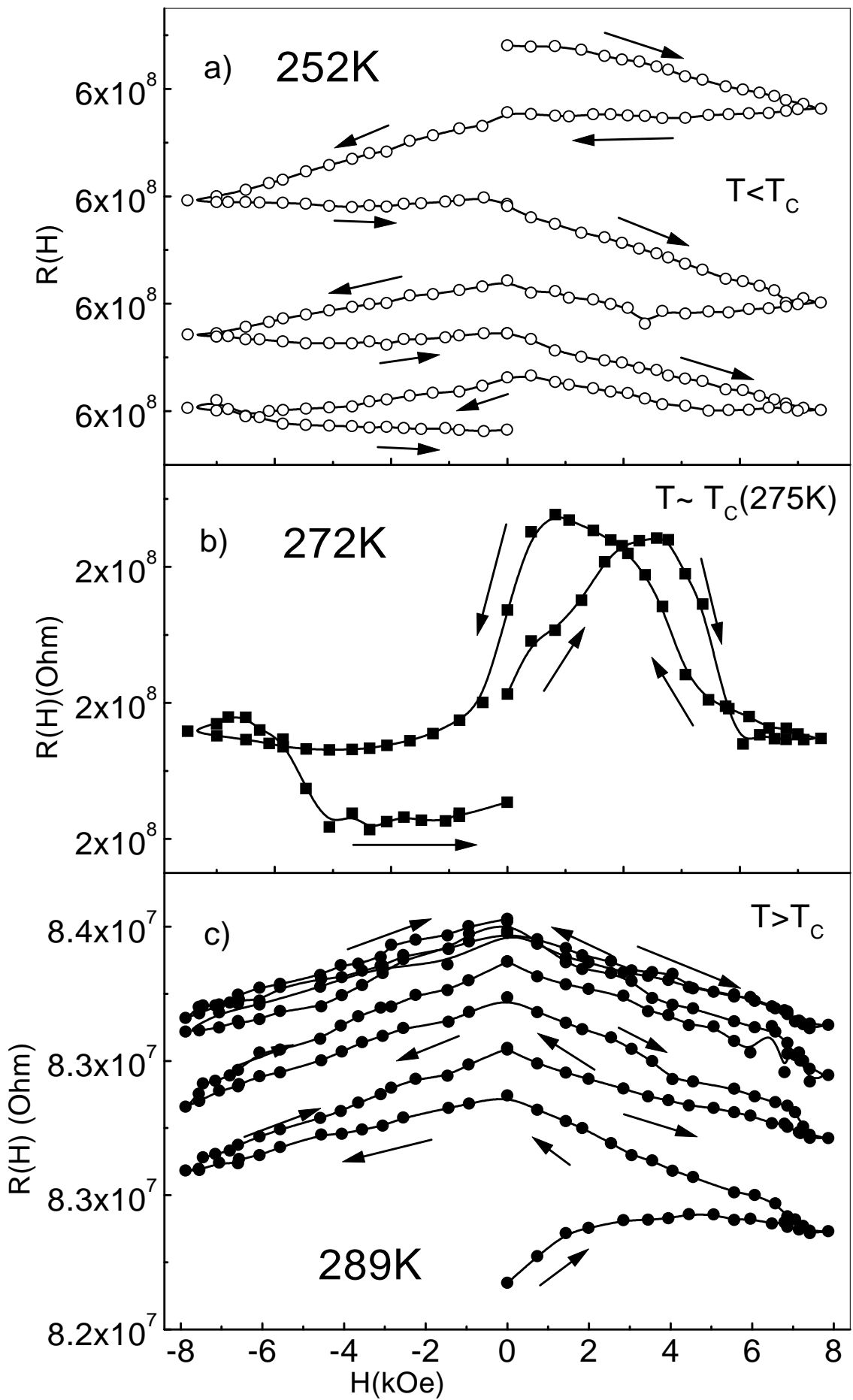
Fig.9 Magnetoresistance measured for more than one cycle of applied field
$\pm$ 8 kOe.
\end{document}